\documentclass[epj]{svjour}

\usepackage{amsmath,amssymb,amsfonts}
\usepackage{graphicx,subfigure}
\usepackage{cite}
\usepackage{epsfig,rotate}
\renewcommand{\l}{\left}
\renewcommand{\r}{\right}
\newcommand{\ii}{\text{i}}

\begin{document}

\title{Thermodynamics of rotating self-gravitating systems}


\author{E.V. Votyakov, A. De Martino \and D.H.E. Gross}

%
%

\institute{Hahn-Meitner-Institut, Bereich Theoretische Physik,
Glienickerstr. 100, 14109 Berlin (Germany)\\
\email{votyakov@hmi.de, demartino@hmi.de, gross@hmi.de}}
\date{}
%
\abstract{\rm We investigate the statistical equilibrium
properties of a system of classical particles interacting via
Newtonian gravity, enclosed in a three-dimensional spherical
volume. Within a mean-field approximation, we derive an equation
for the density profiles maximizing the microcanonical entropy and
solve it numerically. At low angular momenta, i.e. for a slowly
rotating system, the well-known gravitational collapse
``transition'' is recovered. At higher angular momenta, instead,
rotational symmetry can spontaneously break down giving rise to
more complex equilibrium configurations, such as double-clusters
(``double stars''). We analyze the thermodynamics of the system
and the stability of the different equilibrium configurations
against rotational symmetry breaking, and provide the global phase
diagram.  
} 
\maketitle

\section{Introduction}

In this article, we study the equilibrium properties of a system
of $N$ classical particles subject to mutual gravitation, assuming
that this self-gravitating gas is enclosed in a finite
three-dimensional spherical box and rotates around its center. The
Hamiltonian reads
\begin{gather} H_N\equiv
H_N(\{\boldsymbol{r}_i\},\{\boldsymbol{p}_i\})=\frac{1}{2m}
\sum_{i=1}^N p_i^2+\Phi(\{\boldsymbol{r}_i\})\label{ham}\\
\Phi(\{\boldsymbol{r}_i\})=-Gm^2\sum_{1\leq i<j\leq N}
\frac{1}{|\boldsymbol{r}_i-\boldsymbol{r}_j|}\label{pot}
\end{gather}
where $\boldsymbol{r}_i\in V\subset\mathbb{R}^3$,
$\boldsymbol{p}_i\in\mathbb{R}^3$ and $m>0$ denote, respectively,
the position, momentum and mass of the $i$-th particle, while $G$
is the gravitational constant. (In the following, we set $m=1$.)
$V$ stands for the (volume of the) box containing the particles.
Notice that the total potential energy scales as $N^2$. One wants
\begin{enumerate}
\item[a.]to calculate the spatial distribution of particles at
equilibrium, i.e. the most probable microscopic state;
\item[b.]to derive the global phase diagram of the system;
\item[c.]to study its thermodynamics (e.g. caloric curves);
\item[d.]to describe the phase transitions that eventually take
place.
\end{enumerate}

In general, the physics of systems whose microscopic constituents
interact via long-range potentials\footnote{By which we will mean
decaying with the interparticle distance $r$ more slowly than
$r^{-D-\epsilon}$ with $\epsilon>0$ in $D$ dimensions when
$r\to\infty$.}, such as (\ref{ham},\ref{pot}), is highly
non-standard and the analysis of their equilibrium state
represents a considerable theoretical challenge \cite{landau96}.
At odds with more conventional systems with short-ranged forces,
they are not additive (i.e. they cannot be divided into
macroscopic subsystems with negligible mutual interaction) and not
extensive (i.e. the densities of thermodynamic functionals are not
bounded in the limit $N\to\infty$). These facts have several
important consequences. Long-range systems can have negative
specific heat
\cite{lyndenbell68,thirring70,gross82,gross158,lyndenbell99};
statistical ensembles can be inequivalent
\cite{gross158,gross124,gross140,gross174,barre01}\footnote{In
fact, the paper \cite{gross158} pointed explicitly to the failure
of the canonical ensemble near first-order phase transitions in
general, and to its non-equivalence with the fundamental
microcanonical ensemble, which displays a negative heat capacity
there.}; they do not possess a proper infinite-volume limit
\cite{gallavotti99}; and they can attain inhomogeneous
configurations (indeed, even their ground state is inhomogeneous,
and it has been suggested that fractal structures may as well
arise \cite{devega96,devega98}). Conventional statistical
mechanics techniques that apply to homogeneous, short-range
systems hence fail for long-range systems. The key theoretical
problem here is a very fundamental one: to devise a mathematical
framework that allows the study of phase transitions and other
collective phenomena \cite{gross186}.

Self-gravitating gases are the most prominent example of a
long-range system and have attracted a great deal of attention
over the years. Most static theories rely on the microcanonical
ensemble, where conserved macroscopic observables represent the
natural control parameters (see \cite{padmanabhan90} for a
review). The task in this setting is that of finding the most
probable (entropy-maximizing) equilibrium configuration of
(\ref{ham},\ref{pot}) with $V$ finite as a function of the
integrals of (Hamiltonian) motion, the simplest and physically
most relevant being the total energy $E$ and the total angular
momentum
$\boldsymbol{L}=\sum_{i=1}^N\boldsymbol{r}_i\times\boldsymbol{p}_i$,
whose conservation is related to the invariance of (\ref{ham})
under rotations. Taking the conservation of $\boldsymbol{L}$ into
account leads however to enormous technical difficulties. Most
authors have thus neglected rotation by breaking the rotational
symmetry explicitly from the outset, e.g. by taking a
non-spherical $V$, and used $E$ as the only control parameter (see
e.g. \cite{chavanis01,cerruti01,chavanis02} for some recent work).
The analysis of the equilibrium state of self-gravitating gases is
hence a rather well-studied problem without rotation. It turns out
that at high energy (i.e. temperature), where the kinetic term
dominates, the system is most likely found in a homogeneous cloud
(shortly, ``gas'' state) filling the available volume. At low
energy, instead, where the gravitational energy dominates, a
collapsed configuration is preferred, where particles form a
single dense globular cluster lying in an almost void background
(``single star''). This is the well-known gravitational collapse
transition first described in \cite{antonov62}. In between these
two ``phases'' (not being homogeneous, the single-cluster is not a
proper thermodynamic phase), for a whole range of energies, the
specific heat is negative. In this transition regime the canonical
ensemble fails\footnote{It has been recently shown that ensemble
equivalence is restored in the ``dilute'' limit $(N,V)\to\infty$
with $N/V^{1/3}$ fixed, in which thermodynamic functionals exist
\cite{devega02}.}.

For a rotating system, the situation is expected to be
substantially more complex. From a qualitative viewpoint, the
equilibrium density profile will depend on the ratio between the
rotational and gravitational contributions to the total energy.
When the latter dominates, gravitational attraction should cause
the system to collapse. At high ratios, instead, when rotation is
sufficiently fast (high angular momentum), more complex
distributions should arise. Dynamical studies based on
fluid-mechanics techniques
\cite{lyndenbell67,chandrasekhar69,burkert93} suggest that
ring-like and disk-like structures might appear. Ultimately, at
sufficiently high rotational energies, two distinct dense clusters
(i.e. a ``double star'') should form.

The richness suggested by the fluid-dynamical picture cannot be
recovered in a static equilibrium theory without the inclusion of
rotation. Double-cluster configurations can arise in a static
framework only from the spontaneous breaking of the rotational
symmetry of (\ref{ham}), which should take place when the angular
momentum is sufficiently high. Effects connected to rotation
should also lead to the formation of rings and other types of
structures. Despite some attempts \cite{gross181,laliena98},
however, a detailed static theory embodying angular momentum is
lacking.

The purpose of this work, which builds on
\cite{lyndenbell67,laliena98}, is to include angular momentum in
the microcanonical theory. Using a mean-field approximation, we
derive an integral equation for the density profiles corresponding
to stationary points of the microcanonical entropy surface and
solve it numerically as a function of $E$ and
$L=|\boldsymbol{L}|$. The usual collapse transition is recovered
at low angular momentum. At high angular momenta, instead, we find
that a spontaneous breakdown of the rotational symmetry occurs at
sufficiently low energies. This gives rise to more complex
equilibrium structures, including ``double stars'', rings and
disks. We derive the global phase diagram of the self-gravitating
gas in the $(E,L)$ plane. By studying the Hessian of the
microcanonical entropy \cite{gross174,gross173}, we characterize
three pure phases (where the system is found in a ``gas'',
``single star'' and ``double star'' configuration, respectively)
and a large mixed phase with negative specific heat, phase
separation and competition between different equilibrium density
profiles. Finally, we analyze the thermodynamics of the system,
deriving the caloric curves (temperature versus energy) in the
different phases, and analyze the stability of the stationary
distributions.

This paper, which follows \cite{prl}, is structured as follows. In
Sec. 2 we expose the microcanonical mean field theory of
(\ref{ham}), derive the entropy functional and the stationarity
condition. Sec. 3 is dedicated to the results. We report the
numerical solution of the main equation, together with the global
phase diagram and a few equilibrium configurations. Then we pass
to the thermodynamics of the system, with special emphasis on the
rotational-symmetry-breaking transition, the physics of mixed
phase, and the stability problem. Finally, Sec. 4 contains our
conclusions and a some remarks about the work presented here, and
a list of open problems.

\section{Microcanonical mean-field theory}

We consider the system with Hamiltonian as given in
(\ref{ham},\ref{pot}), enclosed in a three-dimensional spherical
volume $V$, to preserve rotational symmetry and ensure
conservation of the total angular momentum $\boldsymbol{L}$. At
the same time, the box breaks translational invariance, hence the
total linear momentum is not conserved. The aim of the
microcanonical theory is to find the particles' density profiles
$\rho(\boldsymbol{r})$ satisfying
\begin{equation}
\int_V\rho(\boldsymbol{r})d\boldsymbol{r}=N
\end{equation}
that maximize the entropy ($k=1$)
\begin{equation}\label{duino}
S_N(E,\boldsymbol{L})=\ln W_N(E,\boldsymbol{L})
\end{equation}
$W_N$ being the microcanonical ``partition sum'' ($h=1$)
\begin{equation}\label{wu}
W_N(E,\boldsymbol{L})=\frac{\epsilon}{N!}\int \delta(H_N-E)
~\delta(\boldsymbol{L}-\sum_{i=1}^N\boldsymbol{r}_i
\times\boldsymbol{p}_i)~D\boldsymbol{r}~D\boldsymbol{p}
\end{equation}
We used the shorthand notations
$D\boldsymbol{r}=\prod_{i=1}^Nd\boldsymbol{r}_i$ and
$D\boldsymbol{p}=\prod_{i=1}^N d\boldsymbol{p}_i$. $\epsilon$ is a
constant that makes $W_N$ dimensionless. Integrals over momenta
are from $-\infty$ to $+\infty$, while integrals over
$\{\boldsymbol{r}_i\}$ are performed over $V^N$. Clearly, such
equilibrium profiles will depend on $E$ and $\boldsymbol{L}$.

We now calculate the microcanonical partition sum (\ref{wu}). To
perform the integrals over momenta, namely to evaluate
\begin{multline}
F_N(\{\boldsymbol{r}_i\},K,\boldsymbol{L})=\\=\int
\delta(K-\frac{1}{2}\sum_{i=1}^N
p_i^2)~\delta(\boldsymbol{L}-\sum_{i=1}^N\boldsymbol{r}_i
\times\boldsymbol{p}_i)~D\boldsymbol{p}
\end{multline}
one can follow Laliena \cite{laliena98} (see also \cite{gross174})
and use the Laplace transform of $F_N$ in $K$, that is ($\Re s>0$)
\begin{equation}
\widetilde{F}_N(\{\boldsymbol{r}_i\},s,\boldsymbol{L})=
\int_0^\infty F_N(\{\boldsymbol{r}_i\},K,\boldsymbol{L})
~e^{-sK}~dK
\end{equation}
Inserting the integral representation of the $\delta$ function and
performing the trivial integral over $K$ one gets
\begin{equation}
\widetilde{F}_N(\{\boldsymbol{r}_i\},s,\boldsymbol{L})=\int e^{\ii
\boldsymbol{\omega\cdot L}-\ii \sum_i\boldsymbol{\omega\cdot
r}_i\times\boldsymbol{p}_i-\frac{s}{2}\sum_i
p_i^2}~D\boldsymbol{p}~D\boldsymbol{\omega}
\end{equation}
where $D\boldsymbol{\omega}=d\boldsymbol{\omega}/(2\pi)^3$. The
above integrals are at most of Gaussian type and can be performed
to yield
\begin{equation}\label{zzzzz}
\widetilde{F}_N(\{\boldsymbol{r}_i\},s,\boldsymbol{L})=
\frac{(2\pi)^{\frac{3N-3}{2}}}{\sqrt{\det\mathbb{I}}}\frac{e^{-\frac{1}{2}s\boldsymbol{L}^T
\mathbb{I}^{-1}\boldsymbol{L}}}{s^{\frac{3N-3}{2}}}
\end{equation}
where $\mathbb{I}\equiv\mathbb{I}(\{\boldsymbol{r}_i\})$ denotes
the inertia tensor, with elements ($a,b=1,2,3$)
\begin{equation}\label{inert}
I_{ab}(\{\boldsymbol{r}_i\})=\sum_{i=1}^N
(r_i^2\delta_{ab}-r_{i,a} r_{i,b})
\end{equation}
and $\boldsymbol{L}^T \mathbb{I}^{-1}\boldsymbol{L}=\sum_{a,b}L_a
I^{-1}_{ab}L_b$. The inverse Laplace transform of (\ref{zzzzz}) is
given by
\begin{equation}
F_N(\{\boldsymbol{r}_i\},K,\boldsymbol{L})=
\frac{(2\pi)^{\frac{3N-3}{2}}}{\Gamma(\frac{3N-3}{2})\sqrt{\det\mathbb{I}}}(K-\frac{1}{2}\boldsymbol{L}^T
\mathbb{I}^{-1}\boldsymbol{L})^{\frac{3N-5}{2}}
\end{equation}
for $K>\frac{1}{2}\boldsymbol{L}^T \mathbb{I}^{-1}\boldsymbol{L}$
and $F_N(\{\boldsymbol{r}_i\},K,\boldsymbol{L})=0$ otherwise.
Hence after integrating out the momenta the microcanonical
partition function reads
\begin{equation}\label{wu2}
W_N(E,\boldsymbol{L})=\frac{\epsilon
A}{N!}\int\frac{[E-\frac{1}{2}\boldsymbol{L}^T
\mathbb{I}^{-1}\boldsymbol{L}-\Phi(\{\boldsymbol{r}_i\})]^{\frac{3N-5}{2}}
}{\sqrt{\det\mathbb{I}}} ~D\boldsymbol{r}
\end{equation}
where $A=(2\pi)^{\frac{3N-3}{2}}/\Gamma((3N-3)/2)$. The term in
square brackets in (\ref{wu2}) is nothing but the kinetic energy
of the system. Now setting for simplicity
$\mathcal{K}=E-\frac{1}{2}\boldsymbol{L}^T
\mathbb{I}^{-1}\boldsymbol{L}-\Phi(\{\boldsymbol{r}_i\})$, we
remark that the integrand is
\begin{equation}
e^{N\l[\frac{3}{2}\log\mathcal{K}-\frac{5}{2N}\log\mathcal{K}-
\frac{1}{2N}\log\sqrt{\det\mathbb{I}}\r]}
\end{equation}
Being interested in the behaviour of the system for large $N$ (see
below), we shall retain just the leading order in $N$ in the above
expression. We are thus left with
\begin{equation}\label{wu3}
W_N(E,\boldsymbol{L})=\frac{\epsilon
A}{N!}\int[E-\frac{1}{2}\boldsymbol{L}^T
\mathbb{I}^{-1}\boldsymbol{L}-\Phi(\{\boldsymbol{r}_i\})]^{\frac{3N}{2}}
~D\boldsymbol{r}
\end{equation}
and it remains to integrate over $V^N$.

To this aim, we write the potential $\Phi$ and the components of
the inertia tensor as functionals of the density profile $\rho$ as
follows:
\begin{gather}
\Phi(\{\boldsymbol{r}_i\})~\to~
\Phi[\rho]=-\frac{G}{2}\int\frac{\rho(\boldsymbol{r})\rho(\boldsymbol{r'})}{
|\boldsymbol{r}-\boldsymbol{r'}|}d\boldsymbol{r}d\boldsymbol{r'}\label{potmf}\\
I_{ab}(\{\boldsymbol{r}_i\})~\to~ I_{ab}[\rho]=\int
\rho(\boldsymbol{r})\l(r^2\delta_{ab}-r_a r_b\r)d\boldsymbol{r}
\end{gather}
Notice that in this way two- and many-body correlations are
neglected. This allows to recast (\ref{wu3}) in the form of the
functional-integral
\begin{equation}\label{wu3}
W_N^{\text{mf}}(E,\boldsymbol{L})\!=\!\frac{\epsilon A
}{N!}\int\![E-\frac{1}{2}\boldsymbol{L}^T
\mathbb{I}^{-1}\boldsymbol{L}-\Phi[\rho]]^{\frac{3N}{2}}
P[\rho]d\rho(\boldsymbol{r})
\end{equation}
(mf $=$ mean field) where $P[\rho]$ is the probability to observe
a density profile $\rho\equiv \rho(\boldsymbol{r})$. To estimate
the latter, we follow the logic of Lynden-Bell
\cite{lyndenbell67}. We subdivide the spherical volume $V$ into
$K$ identical cells labeled by the positions of their centers. The
idea is to replace the integral over $V^N$ with a sum over the
cells. In order to avoid configurations with high densities where
other physical processes (e.g. nuclear reactions) become more
important than gravity, and cure the short-distance singularity of
the Newtonian potential, we assume that each cell may host up to
$n_0$ particles ($1\ll n_0\ll N$). This condition is essentially
equivalent to considering hard spheres instead of point particles.
$P[\rho]$ is now proportional to the number of ways in which our
$N$ particles can be distributed inside the $K$ cells with maximal
capacity $n_0$. Denoting by $n(\boldsymbol{r}_k)$ the number of
particles located inside the $k$-th cell, a simple combinatorial
reasoning leads to
\begin{eqnarray}
P[\rho] & \propto & \frac{N!}{n(\boldsymbol{r}_1)!\cdots
n(\boldsymbol{r}_K)!}\prod_{{\rm cells~}k}
\frac{n_0!}{(n_0-n(\boldsymbol{r}_k))!}\nonumber\\ & = &
N!\prod_{{\rm cells~} k}{n_0\choose n(\boldsymbol{r}_k)}
\end{eqnarray}
where it is understood that the product involves configurations
$\{n(\boldsymbol{r}_k)\}$ such that $\sum_k
n(\boldsymbol{r}_k)=N$. Introducing the relative cell occupancy
\begin{equation}
c(\boldsymbol{r})=\frac{n(\boldsymbol{r})}{n_0}=\frac{V
\rho(\boldsymbol{r})}{K n_0}
\end{equation}
and approximating the factorials by means of Stirling's formula
(assuming $n(\boldsymbol{r}_k)\gg 1$ and
$n_0-n(\boldsymbol{r}_k)\gg 1$), we get
\begin{multline}\label{pidici}
P[c] \propto  N!~e^{-\frac{n_0 K}{V}\int \l[c(\boldsymbol{r})\log
c(\boldsymbol{r})+
(1-c(\boldsymbol{r}))\log(1-c(\boldsymbol{r}))\r]
d\boldsymbol{r}}=\\ = N!~ e^{-\frac{N}{\Theta}\int
 \l[c(\boldsymbol{x})\log c(\boldsymbol{x})+
(1-c(\boldsymbol{x}))\log(1-c(\boldsymbol{x}))\r] d\boldsymbol{x}}
\end{multline}
where we introduced the dimensionless variable
$\boldsymbol{x}=\boldsymbol{r}/R$ and defined the average coverage
\begin{equation}
\Theta=\frac{NV}{n_0 K R^3}=\int c(\boldsymbol{x})d\boldsymbol{x}
\end{equation}
It is simple to check that in terms of $c(\boldsymbol{x})$ the
potential and the inertia tensor are respectively given by
\begin{gather}
\Phi[c]=-\frac{G N^2}{2 R\Theta^2}
\int\frac{c(\boldsymbol{x})c(\boldsymbol{x'})
}{|\boldsymbol{x}-\boldsymbol{x'}|}d\boldsymbol{x}d\boldsymbol{x'}\\
I_{ab}[c]=\frac{NR^2}{\Theta}\int
c(\boldsymbol{x})(x^2\delta_{ab}-x_a x_b)d\boldsymbol{x}
\end{gather}
In the following, we shall measure energies in units of
$\frac{GN^2}{R}$ and inertia tensor components in units of $NR^2$
so that we will be dealing with the reduced (dimensionless)
quantities
\begin{gather}
\Phi[c]=-\frac{1}{2 \Theta^2}
\int\frac{c(\boldsymbol{x})c(\boldsymbol{x'})
}{|\boldsymbol{x}-\boldsymbol{x'}|}d\boldsymbol{x}d\boldsymbol{x'}\\
I_{ab}[c]=\frac{1}{\Theta}\int
c(\boldsymbol{x})(x^2\delta_{ab}-x_a x_b)d\boldsymbol{x}
\end{gather}

Plugging (\ref{pidici}) into (\ref{wu3}), one arrives at the
familiar form
\begin{equation}
W_N^{\text{mf}}(E,\boldsymbol{L})\propto\int e^{N S^{\text{mf}}_N
[c]}~dc(\boldsymbol{x})\label{entra}
\end{equation}
where the ``action'' $S^{\text{mf}}_N$ has the following
expression:
\begin{multline}\label{sigma}
S^{\text{mf}}_N[c]=\frac{3}{2}\log\l[E-\frac{1}{2}\boldsymbol{L}^T
\mathbb{I}^{-1}\boldsymbol{L}-\Phi[c]\r]+\\-\frac{1}{\Theta}\int\l[c(\boldsymbol{x})\log
c(\boldsymbol{x})
+(1-c(\boldsymbol{x}))\log(1-c(\boldsymbol{x}))\r]~d\boldsymbol{x}
\end{multline}
Clearly, $\mathbb{I}\equiv\mathbb{I}[c]$. $S^{\text{mf}}_N$ has an
obvious physical interpretation as the sum of the energetic and
combinatorial contribution to the entropy, respectively. For large
$N$, the integral (\ref{entra}) can be computed as usual by the
steepest-descent (Laplace) method. This immediately yields
\begin{equation}
W_N^{\text{mf}}(E,\boldsymbol{L})\simeq \exp
\l[N\max_{c(\boldsymbol{x})}S^{\text{mf}}_N[c]\r]
\end{equation}
hence the ``physical'' value of the entropy density for large $N$
is nothing but the maximum of $S^{\text{mf}}_N$ over the space of
relative cell occupancies $c$.

An elementary variation of $S^{\text{mf}}_N$ with respect to $c$,
the constraint on $\Theta$ being enforced by a Lagrange multiplier
$\mu$ playing the role of a chemical potential, straightforwardly
leads to the stationarity condition
\begin{equation}\label{key}
\log\frac{c(\boldsymbol{x})}{1-c(\boldsymbol{x})}=-
\frac{\beta}{\Theta}U(\boldsymbol{x})+\frac{1}{2}\beta(
\boldsymbol{\omega}\times\boldsymbol{x})^2-\mu
\end{equation}
or, equivalently,
\begin{equation}\label{key2}
c(\boldsymbol{x})=(1+e^{\frac{\beta}{\Theta}U(\boldsymbol{x})-\frac{1}{2}\beta(
\boldsymbol{\omega}\times\boldsymbol{x})^2+\mu})^{-1}
\end{equation}
where $\boldsymbol{\omega}\equiv\boldsymbol{\omega}[c]$ is the
angular velocity (related to the total angular momentum by the
relation $\boldsymbol{L}=\mathbb{I}\boldsymbol{\omega}$), and
$\beta\equiv\beta[c]$ and $U(\boldsymbol{x})$ are respectively
defined as
\begin{gather}
\beta=\frac{3/2}{[E-\frac{1}{2}\boldsymbol{L}^T
\mathbb{I}^{-1}\boldsymbol{L}-\Phi[c]]}\equiv\frac{3}{2\mathcal{K}}\label{beta}\\
U(\boldsymbol{x})=-\int\frac{c(\boldsymbol{x'})}{
|\boldsymbol{x}-\boldsymbol{x'}|}~d\boldsymbol{x'}\equiv 2
\Theta^2\frac{\delta\Phi}{\delta c(\boldsymbol{x})}\label{Uofx}
\end{gather}
One sees that $\beta$ is related to the kinetic energy of the
systems, i.e. to the (inverse) temperature. The essence of the
mean-field approximation is clearly expressed by the fact that
\begin{equation}
\Phi[c]=\frac{1}{2\Theta^2}\int
c(\boldsymbol{x})U(\boldsymbol{x})d\boldsymbol{x}
\end{equation}
Equation (\ref{key}) (or (\ref{key2})) is our central result.
Functions $c^*$ solving (\ref{key}) and being entropy maxima in
the space of $c$'s represent our desired equilibrium distribution
of particles.

The correct way to analyze the problem consists in solving
(\ref{key}) at fixed energy and angular momentum, subsequently
calculating intensive quantities (temperature and angular
velocity). For the sake of simplicity and without any loss of
generality, we now fix the angular momentum to lie parallel to the
$3$-axis, and concentrate on $|\boldsymbol{L}|=L$. We remark at
this point that Lynden-Bell statistics, which is reminiscent of
Fermi-Dirac statistics in real space (i.e. not in phase space),
plays a crucial role. In fact, once overlapping is ruled out, the
Hamiltonian (\ref{ham}) has a well-defined ground state, with
particles collapsed in a core but without coming too close to each
other. If one used Boltzmann statistics and point particles,
instead, the system would have no ground state, since the
potential energy would be unbounded from below. Antonov
catastrophe \cite{antonov62} can be seen as a direct consequence
of this fact. For this reason, Lynden-Bell statistics is probably
more appropriate for self-gravitating systems\footnote{In order to
restore a ground state, however, the Thirring potential
\cite{thirring70}, which mimics Newtonian gravity, has been used
together with Boltzmann statistics (see e.g. \cite{laliena98}).}.
The existence of a ground state ensures that (\ref{key}) will
always have a solution at fixed $E$ and $L$. Of course, there may
be multiple viable solutions at the same $E$ and $L$, each bearing
its entropy. In such a case, the criterion is simply that the
higher the entropy, the more probable the solution.

Upon varying $E$ and $L$, one can explore different regions of the
parameter space and ultimately obtain the global phase diagram in
the whole $(E,L)$ plane. The main effect one expects from the
inclusion of rotation is that, for sufficiently high angular
momenta, upon decreasing the energy from high values,
rotationally-symmetric solutions (e.g. homogeneous clouds) will
become unstable against fluctuations that break rotational
symmetry, and solutions without rotational symmetry (e.g. ``double
stars'') will bifurcate continuously from them.

The main problem at this point is merely technical. One can only
hope to solve (\ref{key}) or (\ref{key2}) by numerical
integration. However, the implicit dependence of
$U(\boldsymbol{x})$ and $\beta$ on $c(\boldsymbol{x})$ via the
three-dimensional integral (\ref{Uofx}) makes this a formidable
task. Similar considerations hold for $\boldsymbol{\omega}$, which
has to be computed from the relation
$\boldsymbol{L}=\mathbb{I}\boldsymbol{\omega}$. To simplify things
and in particular to reduce the dimensionality of the integrals
involved, we pass to spherical coordinates,
$\boldsymbol{x}=(x,\theta,\phi)$, and expand the Newtonian
potential in series of real spherical harmonics (see e.g.
\cite{wyld76}):
\begin{equation}\label{newta}
\frac{1}{|\boldsymbol{x}-\boldsymbol{x'}|}\!=\!\sum_{l=0}^\infty
\!\sum_{m=-l}^l\!\frac{4\pi}{2l+1}\frac{(x\vee x')^l}{(x\wedge
x')^{l+1}}Y_{lm}(\theta,\phi)Y_{lm}(\theta',\phi')
\end{equation}
with $x\vee x'=\min\{x,x'\}$ and $x\wedge x'=\max\{x,x'\}$. At the
same time, we formally expand also the relative occupancy $c$:
\begin{equation}
c(\boldsymbol{x})=\sum_{l=0}^\infty \sum_{m=-l}^l
b_{lm}(x)Y_{lm}(\theta,\phi)\label{ser}
\end{equation}
$b_{lm}(x)$ is a radial function whose precise form we will have
to derive. Using the above series, together with the completeness
relation for our basis set $\{Y_{lm}\}$,
\begin{equation}
\int
Y_{lm}(\theta,\phi)Y_{l'm'}(\theta,\phi)~d\cos\theta~d\phi=\delta_{ll'}\delta_{mm'}
\end{equation}
one can easily show that
\begin{gather}
U(\boldsymbol{x})=\sum_{l,m}u_{lm}(x)Y_{lm}(\theta,\phi)\\
u_{lm}(x)=-\frac{4\pi}{2l+1}\int\frac{(x\vee x')^l}{(x\wedge
x')^{l+1}} b_{lm}(x')(x')^2 dx' \label{u}
\end{gather}
Multiplying both sides of (\ref{key2}) by $Y_{lm}$ and integrating
over angular variables one obtains for $b_{lm}$ the system of
integral equations
\begin{gather}
b_{lm}(x)=\int g(x,\theta,\phi) Y_{lm}(\theta,\phi)~d\cos\theta
~d\phi\label{due}\\ g(x,\theta,\phi)=\l[1+e^{\frac{\beta}{\Theta}
\sum_{l,m} u_{lm}(x)Y_{lm}(\theta,\phi) -\frac{1}{2}\beta\omega^2
x^2 \sin^2\theta+\mu}\r]^{-1}\nonumber
\end{gather}
where $l=0,1,\ldots$ and $m=-l,-l+1,\ldots,l$. Notice that
$u_{lm}$, $\beta$ and $\boldsymbol{\omega}$ depend on $b_{lm}$.
This system is completely equivalent to (\ref{key2}), but at least
an iterative solution procedure is imaginable. After having fixed
$E$ and $L$, starting from an initial reasonable guess for
$b_{lm}(x)$, one can compute $u_{lm}(x)$ from (\ref{u}) ($1$-dim.
integral) (and $\beta$ from (\ref{beta})). Using this, $b_{lm}(x)$
can be re-calculated from (\ref{due}) ($2$-dim. integral) to
improve the guess, and so until convergence via e.g. a simple
Newton-Raphson method. As the starting point, it is convenient to
take a high-energy configuration, where the kinetic contribution
is expected to be much larger than the gravitational one, and a
homogeneous cloud (``gas'') type of $c(\boldsymbol{x})$ is very
likely to solve (\ref{key2}). In particular, one can initiate from
a totally symmetric configuration where $b_{00}(x)=\Theta$ and
$b_{lm}(x)=0$ for $(l,m)\neq (0,0)$.

\begin{figure}
\begin{center}
\subfigure[Fixed $\phi=\pi/8$, variable
$l_{\text{max}}$]{\scalebox{.58}{\includegraphics{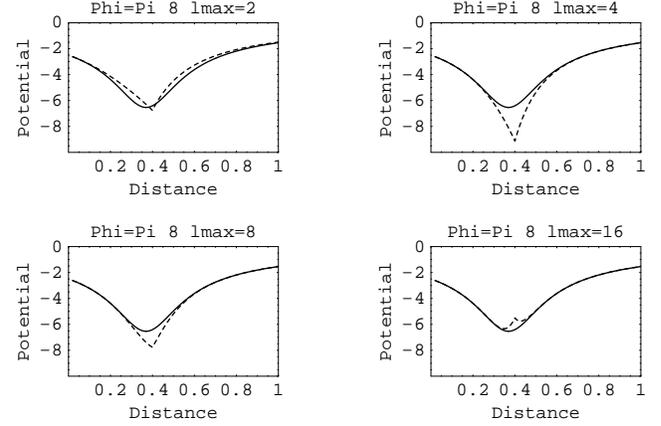}}}
\subfigure[Fixed $l_{\text{max}}=16$, variable
$\phi$]{\scalebox{.58}{\includegraphics{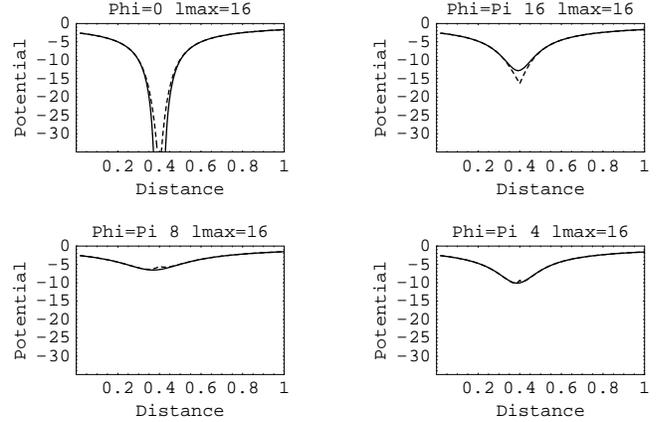}}}
\caption{\label{mox}Effects of truncation of the series
(\ref{newta}) to the term of order $l_{\text{max}}$ (see text for
details). Dashed and continuous lines represent the truncated
potential and the Newtonian potential, respectively.}
\end{center}
\end{figure}

\begin{figure}
\begin{center}
\includegraphics[width=8.5cm]{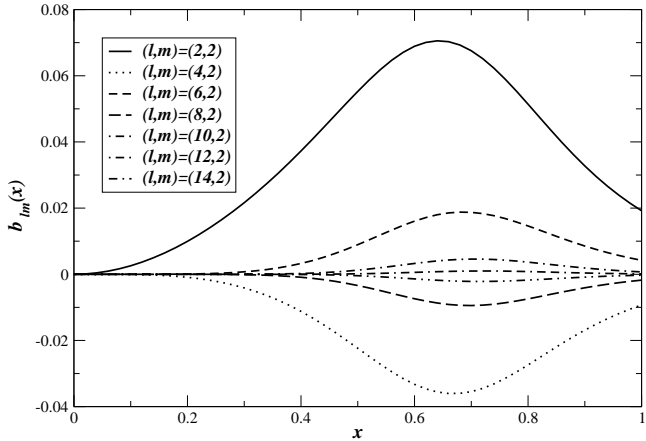}
\caption{\label{spag}Typical behaviour of the radial function
$b_{lm}(x)$ for different $l$ ($l$ even, $2\leq l\leq 14$) at
fixed $m=2$. This particular plot was obtained for $E=-0.18$ and
$L=0.44$.}
\end{center}
\end{figure}

Clearly, actual calculations must be performed with a finite
number of harmonics $l_{\text{max}}$, i.e. the series
(\ref{newta}) must be truncated. The effects of such a truncation
are shown in Fig.~\ref{mox}.
One sees the potential felt by one particle due to another one
fixed at the position $x=0.4$ with $\phi=0$. In (a), the first
particle moves from $x=0$ to $x=1$ at fixed $\phi=\pi/8$ and the
true potential is compared with the truncated one, with maximum
number of included harmonics variable from $2$ to $16$. The latter
case clearly reproduces the Newtonian force with a good degree of
accuracy. However, the $\phi$-dependence must be considered also.
This is shown in (b), where we fixed $l_{\text{max}}=16$ and
measured the potential varying the $\phi$ of the second particle,
keeping the first one fixed. It is clearly seen that the truncated
potential works fine sufficiently far from the ``probe'' particle,
while small deviations occur when the particles are too close.
However this short-distance problem is substantially cured by our
choice to deal with non-overlapping particles. A further hint
about what should be the maximum order of harmonics to be included
in the calculation comes from the study of the behaviour of
$b_{lm}(x)$ for typical solutions of (\ref{due}), an example of
which is reported in Fig.~\ref{spag}.
One clearly sees that $b_{lm}$ dies out as $l$ increases, and that
already for $l=14$ it is for all practical purposes zero.

Hence, we solved (\ref{due}) taking $l_{\text{max}}=16$. We also
excluded odd harmonics. Simple symmetry considerations suggest
that exclusion of $l=1$ harmonics fixes the center of mass in the
origin, while absence of higher-order odd harmonics prevents the
formation of asymmetric structures (e.g. two clusters of different
sizes lying at different distances from the origin). Their effects
on the phase diagram will be studied elsewhere \cite{prep}.
Finally, we measured energy and angular momentum in units of
$GN^2/R$ and $(RGN^3)^{1/2}$, respectively, and took $\Theta=0.02$
always\footnote{The $\Theta$-dependence of the results is an
important issue. In fact, when $\Theta$ is too large even the
usual gravitational collapse transition does not take place
because of particles jamming \cite{padmanabhan90}.}. The results
of this analysis are reported in the next section.

\section{Thermodynamics}

\subsection{Phase diagram}

\begin{figure}
\begin{center}
\includegraphics[width=6.5cm,angle=-90]{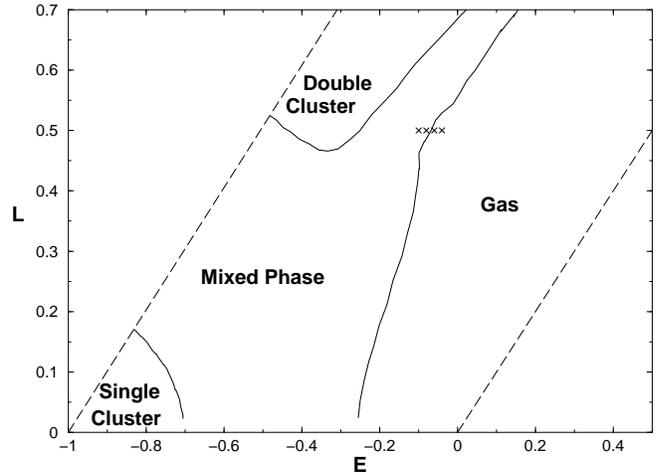}
\caption{\label{phased}Phase diagram in the $(E,L)$-plane. The
dashed lines $E-L=1$ (left) and $E=L$ (right) delimit the region
where the Hessian was calculated. The four markers ($\times$)
correspond to the four situations described in Fig.~\ref{ixxiyy}.}
\end{center}
\end{figure}

\begin{figure}
\begin{center}
\subfigure[$E=-0.72$,
$L=0.4$]{\scalebox{.53}{\includegraphics{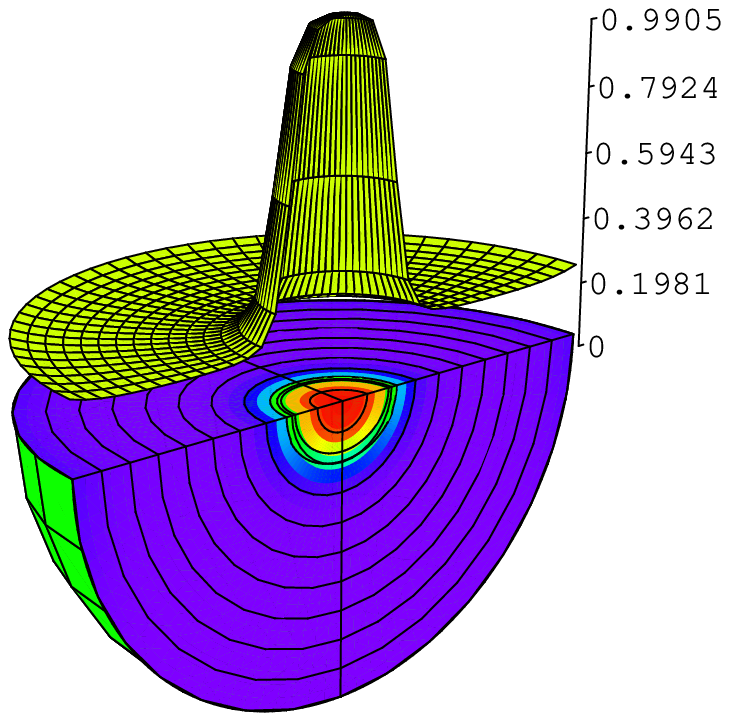}}}
\subfigure[$E=-0.9$,
$L=0.5$]{\scalebox{.53}{\includegraphics{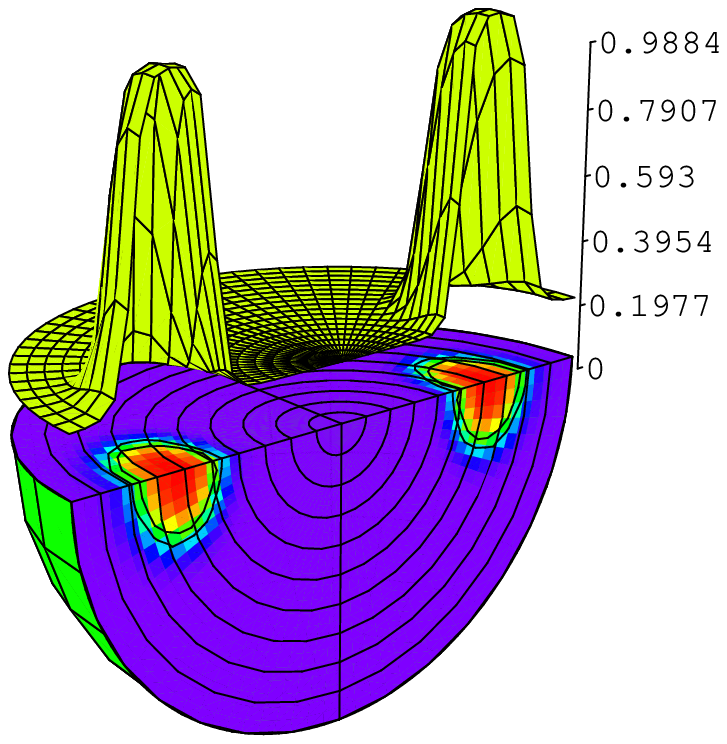}}}
\subfigure[$E=-0.06$,
$L=0.4$]{\scalebox{.53}{\includegraphics{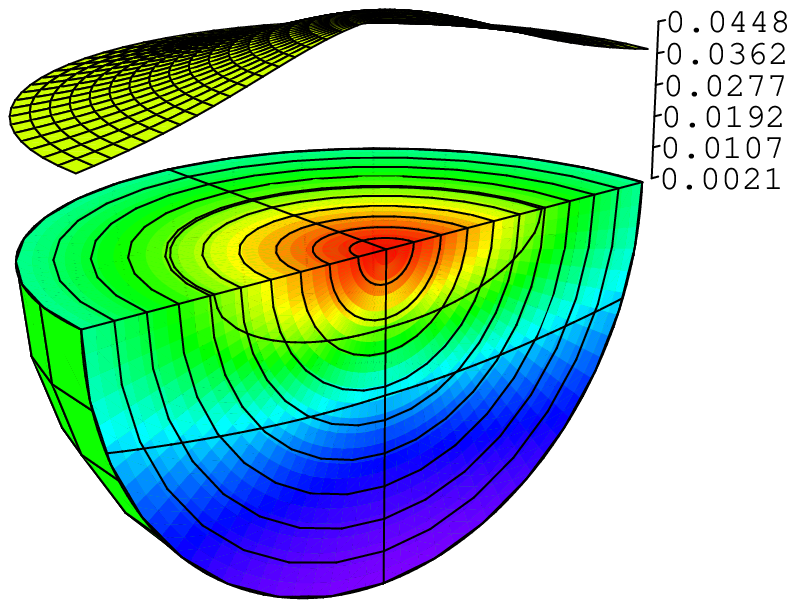}}}
\subfigure[$E=-0.42$,
$L=0.5$]{\scalebox{.53}{\includegraphics{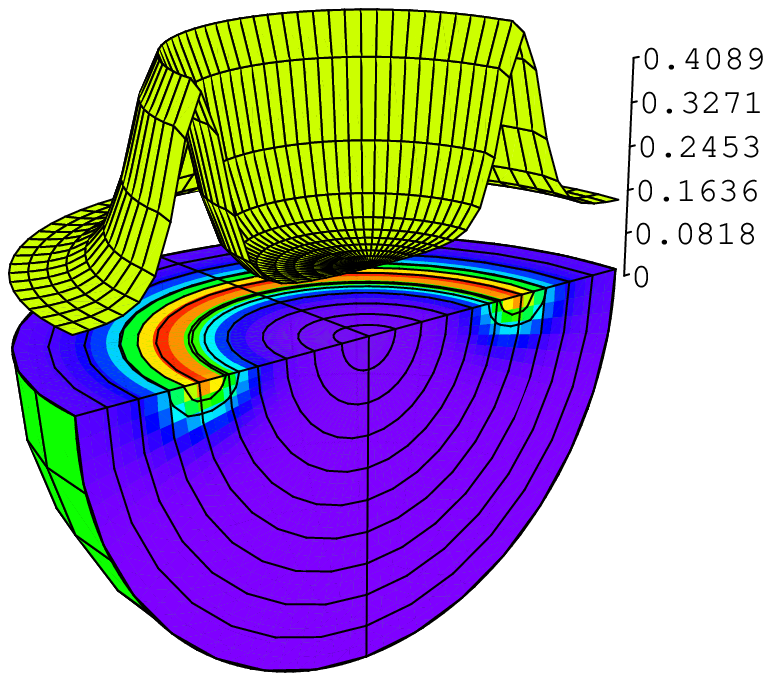}}}
\caption{\label{distri}Examples of stationary distributions
$c(\boldsymbol{x})$ occurring inside our spherical box. Shown are
the contour plot and, above it, the density profile: (a) ``single
star'', (b) ``double star'', (c) ``disk'', (d) ``ring''. }
\end{center}
\end{figure}

We shall discuss here solutions of (\ref{due}) obtained at fixed
$E$ and $L$ in a slab of the $(E,L)$ plane delimited by the lines
$E-L=1$ and $E=L$. The entropy corresponding to each solution can
be calculated via (\ref{sigma}). Pure thermodynamic phases with
one (macroscopic) equilibrium state can be discerned from phase
coexistence regions by studying the Hessian of $S$ in $E$ and $L$,
i.e.
\begin{equation}
{\rm Hes}_{(E,L)}[S]={\rm det}
\begin{pmatrix} \partial^2_E S & \partial_L\partial_E S
\\ \partial_E\partial_L S & \partial_L^2 S
\end{pmatrix}
\end{equation}
In the microcanonical ensemble pure phases are characterized as
having ${\rm Hes}_{(E,L)}[S]>0$ (the entropy as a function of $E$
and $L$ is concave), while phase coexistence regions have ${\rm
Hes}_{(E,L)}[S]<0$ \cite{gross174,gross173} (the entropy has a
convex intruder). Mixed phases are characterized by negative
specific heat and hence ensemble inequivalence. (The reader is
referred to \cite{gross174} for an introductory account on
microcanonical thermostatistics.) Fig.~\ref{phased} shows the
resulting global phase diagram of the system. In Fig.~\ref{distri}
one sees a sample of stationary distributions, together with the
values of $E$ and $L$ at which they were obtained.


Results can be summarized as follows. For low angular momenta:
\begin{enumerate}
\item[a.] at high energies (the kinetic term dominates), the solution
of (\ref{due}) is unique and is of the homogeneous cloud (``gas'')
type; one finds that ${\rm Hes}_{(E,L)}[S]>0$ and the
corresponding (pure) phase is labeled as ``gas'' phase;
\item[b.] at low energies, where gravity dominates, the solution of
(\ref{due}) is unique and of the single-cluster type (``single
star'', e.g. Fig.~\ref{distri}a); correspondingly, one finds a
pure thermodynamic state with ${\rm Hes}_{(E,L)}[S]>0$, which we
call ``single-cluster'' state;
\item[c.] in between these two regimes, one finds a phase coexistence
region with ${\rm Hes}_{(E,L)}[S]<0$ and negative specific heat,
in which different solutions occur at each point $(E,L)$ (``mixed
phase''); here, single-cluster and gas type of solutions occur.
\end{enumerate}
For slowly rotating systems one thus recovers a scenario that is
similar to the usual gravitational collapse that is found in
theories without angular momentum. For higher angular momenta
(i.e. for a rapidly rotating system), instead:
\begin{enumerate}
\item[d.] at high energies, the situation gas-type of solutions
are obtained, and the situation is as in a. above;
\item[e.] at low energies, the Hessian is positive and solutions
of (\ref{due}) are of double-cluster type (``double star'', e.g.
Fig.~\ref{distri}b), and the corresponding phase, labeled ``double
cluster'' is pure;
\item[f.] at intermediate energies, multiple solutions are found,
both of double-cluster type and deformed gas-type. The latter in
particular can be disks (e.g. Fig.~\ref{distri}c) or, if the
angular momentum is high enough, rings (e.g. Fig.~\ref{distri}d).
Correspondingly, the Hessian is negative and we have a phase
coexistence region with negative specific heat.
\end{enumerate}

\begin{figure}
\begin{center}
\includegraphics[width=7cm,angle=-90]{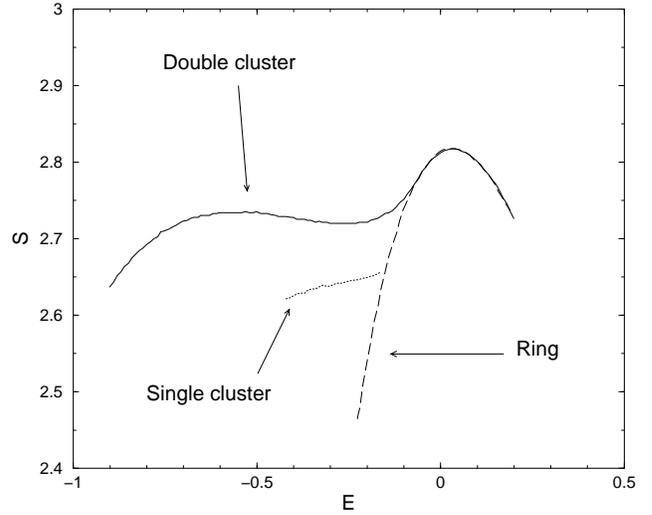}
\caption{\label{trisoluz}Entropy as a function of energy at fixed
$L=0.5$ for three different solutions, as shown. To make the
convex intruder in the entropy (corresponding to the mixed phase)
more evident, we subtracted the quantity $(15/4)E$ from the
entropy.}
\end{center}
\end{figure}

The system thus turns out to have three pure phases (``gas'',
``single star'' and ``double star''), separated by a large mixed
phase. The occurrence of double-star-like solutions is the most
remarkable effect of rotation. To get an idea of the coexistence
of different solutions in the mixed phase, in Fig.~\ref{trisoluz}
we plot the entropy for three different solutions (ring, single
cluster, double cluster) in a range of energy at fixed $L=0.5$.
One sees that the entropy has a ``convex intruder'', corresponding
to the mixed phase and implying negative specific heat. The three
solutions coexist in a whole range of energies, while at low
energies, as evident from the phase diagram, the double-cluster
solution only survives. The point where the rotationally
asymmetric double-cluster solution bifurcates from the
rotationally symmetric ring solution corresponds to the beginning
of the mixed phase at $L=0.5$. In the mixed phase: double-cluster
and (deformed) gas type of configurations compete at high $E$ and
$L$; single-cluster and gas compete at low $L$; finally,
single-cluster and double-cluster compete at intermediate values
of $L$.

The crucial issue of stability (i.e. which of these configurations
are actually entropy maxima in the space of $c$'s) will be dealt
with in Sec. 3.4. For the moment, let it suffice to say that in
the mixed phase, rotationally symmetric structures are unstable to
perturbations that break rotational symmetry. Hence ring
configurations, which also occur in the mixed phase, are not
stable. Deformed gas configurations (e.g. disks or rings) that
occur in the ``gas'' phase, e.g. close to the phase boundary, are
stable. Before analyzing the different transitions that take
place, we shall briefly discuss the important issue of negative
specific heat.

\subsection{Caloric curves, specific heat}

\begin{figure}
\begin{center}
\subfigure[]{\scalebox{.66}{\includegraphics{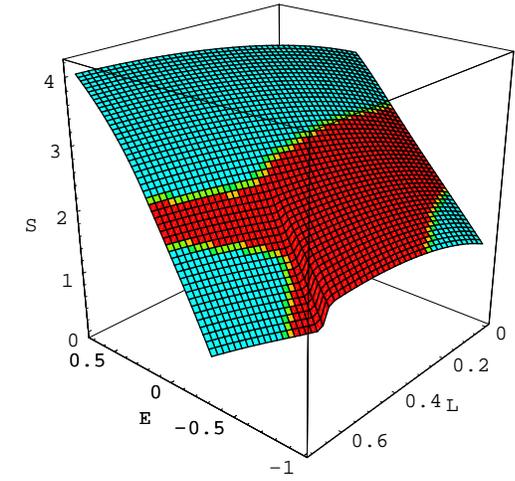}}}
\subfigure[]{\scalebox{.75}{\includegraphics{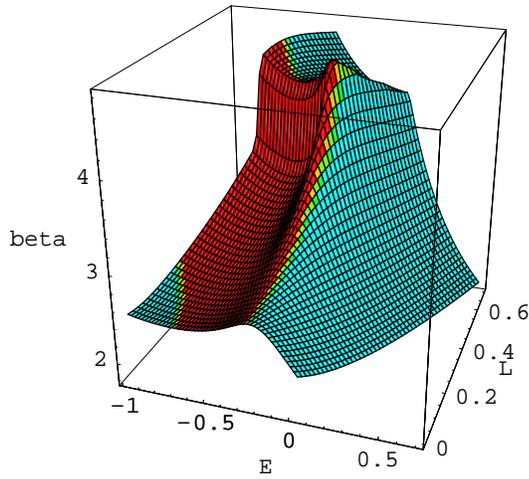}}}
\caption{\label{entropy}(a) Microcanonical entropy $S$ as a
function of $E$ and $L$. In the central region the Hessian ${\rm
Hes}_{(E,L)}[S]$ is negative, signaling the separation of multiple
phases (mixed phase). In the three clear regions, the Hessian is
positive and pure thermodynamic phases (gas, single star, binary
star, respectively) occur. (b) Behaviour of $\beta=\partial_E S$
as a function of $E$ and $L$.}
\end{center}
\end{figure}

\begin{figure}
\begin{center}
\subfigure[$L=0.2$]{\scalebox{.55}{\includegraphics{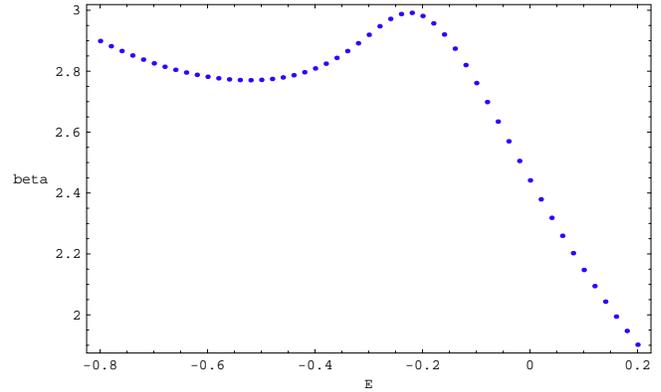}}}
\subfigure[$L=0.4$]{\scalebox{.55}{\includegraphics{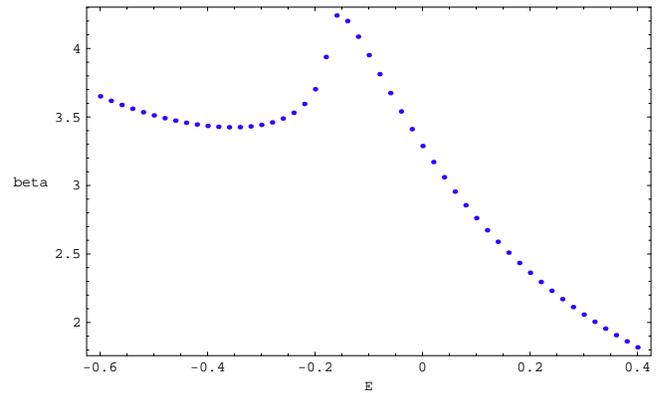}}}
\subfigure[$L=0.6$]{\scalebox{.55}{\includegraphics{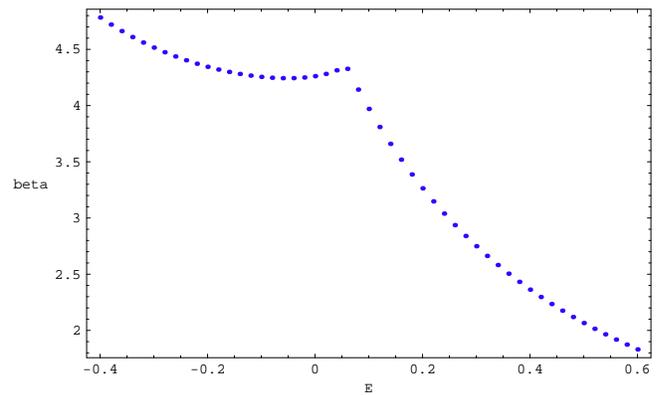}}}
\caption{\label{caloric} Cross sections of the $\beta$-surface
(inverse microcanonical temperature, caloric curves) at different
angular momenta.}
\end{center}
\end{figure}

Once the solutions are obtained, the microcanonical entropy
surface can be immediately calculated from (\ref{sigma}). It is
reported in Fig.~\ref{entropy} together with
the $\beta$-surface, namely
\begin{equation}
\beta\equiv\beta(E,L)=\l(\frac{\partial S}{\partial
E}\r)_{L=\text{constant}}\equiv\frac{1}{T}
\end{equation}
representing the inverse microcanonical temperature as a function
of energy and angular momentum. The central region in the entropy
surface corresponds to the mixed phase and has ${\rm
Hes}_{(E,L)}[S]<0$. Slices of the $\beta$-surface at different $L$
(caloric curves) are shown in Fig.~\ref{caloric}.
One sees that $\beta$ is increasing with $E$ for a whole range of
energies. This means that in that range
$\frac{\partial\beta}{\partial E}>0$, or equivalently that
$\frac{\partial E}{\partial T}<0$, i.e. that the specific heat is
negative. Physically, if the system is heated (increase of total
energy) its temperature diminishes, and vice-versa if it loses
energy its temperature increases. This is well-known to happen in
stars. As they irradiate and release energy, they become hotter
and hotter and contract. Such a behaviour, however, is not
peculiar to self-gravitating systems, but is the generic signal of
a phase separation in the microcanonical ensemble of finite
systems \cite{gross174,gross186} and is not recoverable in the
canonical ensemble, where the specific heat is proportional to
energy fluctuations, i.e. non-negative definite.

\subsection{Phase transitions, symmetry breaking}

\begin{figure}
\begin{center}
\subfigure[$L=0.2$]{\scalebox{.65}{\includegraphics{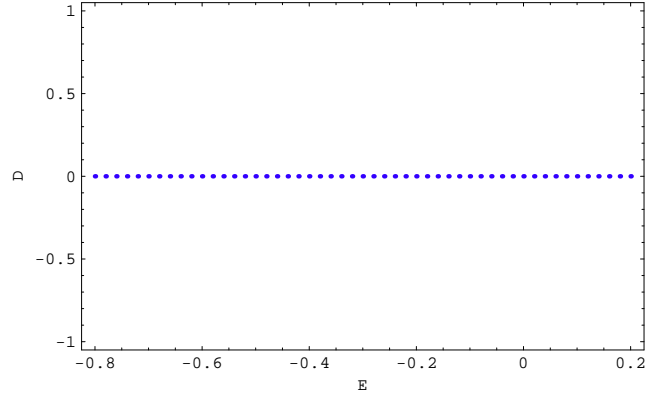}}}
\subfigure[$L=0.6$]{\scalebox{.65}{\includegraphics{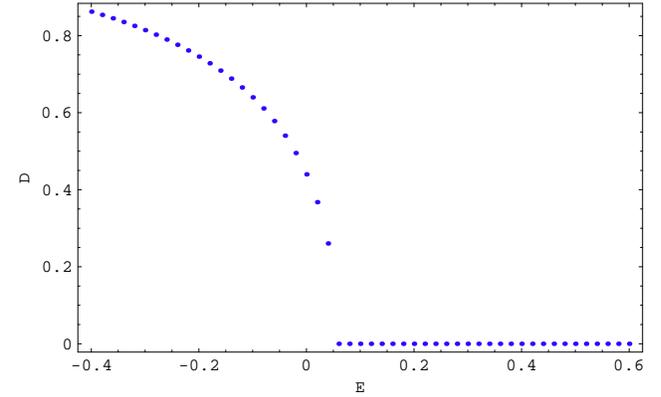}}}
\caption{\label{orderp}Behaviour of the order parameter
$D=I_{11}-I_{22}$ as a function of energy $E$ at fixed angular
momentum $L$. At high $L$, breaking of rotational symmetry is
signaled by a $D\neq 0$.}
\end{center}
\end{figure}

We now turn to investigating the phase transitions. The
low-angular-momentum collapse transitions are analogous to those
discussed at length in the literature. A probably convenient order
parameter to describe them is the density contrast, defined as the
center-to-edge ratio of the particle density. We shall concentrate
here on the rotational-symmetry-breaking transition to
double-cluster solutions, which is the truly novel phenomenon
introduced by rotation. A convenient order parameter to detect
such a transition is
\begin{equation} \label{d}
D=|I_{11}-I_{22}|
\end{equation}
that is, the difference between $1$ and $2$ diagonal components of
the inertia tensor (\ref{inert}). One expects $D$ to be zero when
the solution is rotationally-symmetric, and non-zero for a
solution without rotational symmetry. The reason is physically
clear. If $L=0$ (i.e. in absence of rotation) the system is
necessarily isotropic ($I_{11}=I_{22}=I_{33}$) and rotational
symmetry cannot be broken. When $L\neq 0$, anisotropies may occur
($I_{33}\neq I_{11},I_{22}$) and one can have either
rotationally-homogeneous ($I_{11}=I_{22}$) or
rotationally-heterogeneous ($I_{11}\neq I_{22}$) solutions. The
latter correspond to double clusters. (We remind the reader that
the angular momentum is chosen to lie parallel to the $3$-axis).
In Fig.~\ref{orderp} we show explicitly that $D$ actually behaves
as a conventional order parameter for the
rotational-symmetry-breaking transition (and the appearance of
double clusters) by plotting it as a function of $E$ at low and
high angular momentum. By comparing Fig.~\ref{orderp}b with the
phase diagram the reader can notice that the transition occurs
exactly at the phase boundary.

\begin{figure}
\begin{center}
\includegraphics[width=8.5cm]{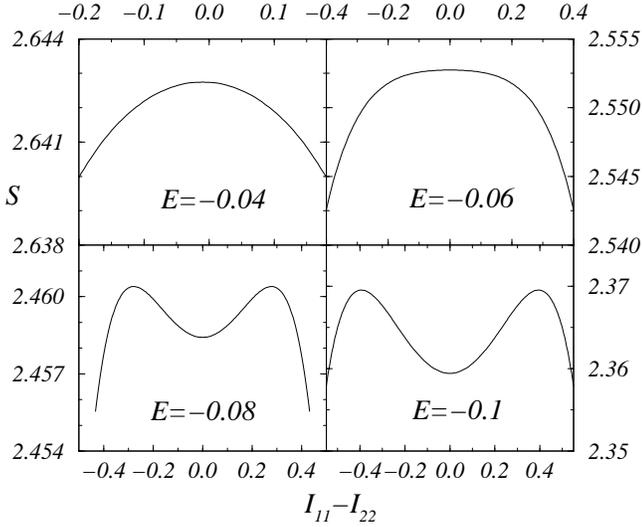}
\caption{\label{ixxiyy}Entropy as a function of $I_{11}-I_{22}$ at
$L=0.5$ and different values of $E$. The values of $E$ and $L$ for
the four figures correspond to the four markers ($\times$) shown
in Fig.~\ref{phased}.}
\end{center}
\end{figure}

Another view of the same transition is given in Fig.~\ref{ixxiyy}.
One sees the microcanonical entropy as a function of
$I_{11}-I_{22}$ at fixed $L=0.5$ and different energies. The four
diagrams shown correspond to the four markers displayed across the
gas-double cluster phase boundary in the phase diagram. The
entropy is clearly seen to develop two peaks at non-zero values of
$I_{11}-I_{22}$, corresponding to double clusters systems, with
the two stars either aligned on the $1$-axis or on the $2$-axis,
respectively. The fact that $S$ becomes flat at the phase
transition indicates that the transition is second-order. The
minimum of $S$, occurring at $I_{11}=I_{22}$, corresponds to
another, rotationally-symmetric solution of (\ref{key2}). In
particular, it is a ring. This brings us to the problem of
stability and clarifies further the structure of the mixed phase:
at fixed $E$ and $L$, the entropy in the $c$ space has (at least)
two maxima corresponding to double stars aligned on different
axes. These are the only stable configurations.

\subsection{Stability}

Usually, the analysis of the (local) stability properties of the
stationary points of the microcanonical entropy (\ref{sigma}) at
fixed mass, energy and angular momentum relies on the study the
sign of second variation of the entropy. In the reference frame of
the principal inertia axes, where $\mathbb{I}$ is diagonal, one
can calculate such a variation explicitly. Omitting details of the
lengthy calculation, one finds
\begin{multline}
\delta^2 S=\frac{\beta}{\Theta^2}\int\frac{\delta
c(\boldsymbol{x}) \delta c
(\boldsymbol{x}')}{|\boldsymbol{x}-\boldsymbol{x}'|}d\boldsymbol{x}
d\boldsymbol{x}'-\frac{1}{\Theta} \int\frac{(\delta c
(\boldsymbol{x}))^2}{c(\boldsymbol{x})(1-c(\boldsymbol{x}))}d\boldsymbol{x}+\\
-\frac{2}{3\Theta^2}\l[\int \delta c (\boldsymbol{x})\log\frac{c
(\boldsymbol{x})}{1-c(\boldsymbol{x})}d\boldsymbol{x}\r]^2+\\
-\beta\sum_{a=1}^3\frac{1}{I_{aa}}\l[\int \delta c
(\boldsymbol{x})\boldsymbol{\omega}^T\l(\frac{\delta\mathbb{I}^{(a)}}{\delta
c(\boldsymbol{x})}\r)\r]^2
\end{multline}
with $\delta c(\boldsymbol{x})$ a mass-preserving perturbation
($\int \delta c(\boldsymbol{x})d\boldsymbol{x}=0$).
$\mathbb{I}^{(a)}\equiv\mathbb{I}^{(a)}[c]$ stands for the $a$-th
column of the inertia tensor $\mathbb{I}$. It is easy to show that
for $\beta\to 0$, that is at sufficiently high energy where the
kinetic term dominates, homogeneous gas-like stationary
configurations are stable against any perturbation. It suffices to
observe that when $\beta\to 0$ the above equation reduces to
($\kappa>0$ constant)
\begin{equation}
\delta^2 S(\beta\to 0)=-\kappa\int(\delta c(\boldsymbol{x}))^2
d\boldsymbol{x}<0
\end{equation}
because $c(\boldsymbol{x})=\text{constant}$ and $\delta
c(\boldsymbol{x})$ is mass-preserving by assumption.

In a more general setting, the second variation of the entropy can
be written as the quadratic form
\begin{equation}
\delta^2 S=\int f(\boldsymbol{x})K(\boldsymbol{x},\boldsymbol{x'})
f(\boldsymbol{x'})~d\boldsymbol{x}~d\boldsymbol{x'}
\end{equation}
where the kernel $K$ is given by
\begin{multline}\label{ka}
K(\boldsymbol{x},\boldsymbol{x}')=-\beta\frac{\delta^2\Phi}{\delta
c(\boldsymbol{x})\delta
c(\boldsymbol{x}')}-\frac{1}{\Theta}\frac{\delta(\boldsymbol{x}-
\boldsymbol{x}')}{c(\boldsymbol{x})(1-c(\boldsymbol{x}')}+\\
-\frac{2\beta^2}{3}\l[\frac{\delta\Phi}{\delta
c(\boldsymbol{x})}-\frac{1}{2}\boldsymbol{\omega}^T\frac{\delta\mathbb{I}}{\delta
c(\boldsymbol{x})}\boldsymbol{\omega}\r]
\l[\frac{\delta\Phi}{\delta
c(\boldsymbol{x}')}-\frac{1}{2}\boldsymbol{\omega}^T\frac{\delta\mathbb{I}}{\delta
c(\boldsymbol{x}')}\boldsymbol{\omega}\r]+\\
-\frac{\beta}{2}\boldsymbol{\omega}^T\l[\frac{\delta\mathbb{I}}{\delta
c(\boldsymbol{x}')}\mathbb{I}^{-1}\frac{\delta\mathbb{I}}{\delta
c(\boldsymbol{x})}+\frac{\delta\mathbb{I}}{\delta
c(\boldsymbol{x})}\mathbb{I}^{-1}\frac{\delta\mathbb{I}}{\delta
c(\boldsymbol{x}')}\r]\boldsymbol{\omega}
\end{multline}
and where as before $\Phi$ stands for the Newtonian potential.
Stability analysis is then equivalent to studying the eigenvalue
problem for $K$, namely
\begin{equation}\label{eigen} \int
K(\boldsymbol{x},\boldsymbol{x}')f(\boldsymbol{x}')d\boldsymbol{x}'
= \lambda f(\boldsymbol{x})
\end{equation}
(see e.g. \cite{padmanabhan89}). From a mathematical viewpoint,
this task is extremely sophisticated. We shall therefore limit
ourselves here to discuss the stability of rotationally-symmetric
configurations against perturbations that break rotational
symmetry, deferring the reader to a later publication for a more
complete analysis of the stability problem.

From the analysis of the preceding section, and in particular from
Fig.~\ref{ixxiyy}, it stems that rotationally symmetric
configurations become unstable against perturbations that break
rotational symmetry at sufficiently high angular momenta and
energies. Two rotationally asymmetric solutions of (\ref{due})
bifurcate continuously from the rotationally symmetric state. At
least for what concerns this class of perturbations, it is safe to
claim that rotationally symmetric structures are stable up to the
phase boundary. Among them, one finds gas-like homogeneous
configurations, and deformed-gas configurations (i.e. disks and
rings). At the phase boundary, solutions with $D=0$ become entropy
minima at least along one direction in the $c$ space, and are no
longer stable against rotational symmetry breaking. Instead, the
only stable configurations in the mixed phase at high enough
angular momentum are double-cluster like ($D\neq 0$).

A more technical argument that supports this conclusion is the
following. Let (see (\ref{key2}))
\begin{equation}\label{g}
G[c]=(1+e^{\frac{\beta}{\Theta}U(\boldsymbol{x})-\frac{1}{2}\beta(
\boldsymbol{\omega}\times\boldsymbol{x})^2+\mu})^{-1}-c(\boldsymbol{x})
\end{equation}
It is easy to show \footnote{It is sufficient to note that
\begin{displaymath} \frac{\delta S}{\delta
c(\boldsymbol{x})}=\frac{1}{\Theta}\log\frac{(1-c(\boldsymbol{x}))(c(\boldsymbol{x})
+G[c])}{c(\boldsymbol{x})(1-c(\boldsymbol{x})-G[c])}
\end{displaymath}
Taking the functional derivative of this expression with respect
to $c(\boldsymbol{x'})$ and evaluating it at the stationary point,
one finds (\ref{sgama}) with $\gamma=[\Theta
c(\boldsymbol{x})(1-c(\boldsymbol{x}))]^{-1}$.} that
\begin{equation}\label{sgama}
\l[\frac{\delta^2 S}{\delta c(\boldsymbol{x})\delta
c(\boldsymbol{x'})}\r]_{G[c]=0}=\gamma\l[\frac{\delta G}{\delta
c(\boldsymbol{x'})}\r]_{G[c]=0}
\end{equation}
with $\gamma>0$, i.e. that the second functional derivative of the
entropy evaluated at the stationary point vanishes together with
the functional derivative of $G$ evaluated at the same point, and
that the two have the same sign. This implies that the stability
analysis can be reduced to the study of the sign of $\delta
G/\delta c$. However, the latter problem is dealt with when
applying the Newton-Raphson method to solve (\ref{key2}). Starting
from high energies, in order to provoke a rotationally asymmetric
solution an appropriate external field (in this case, it is
related to the order parameter discussed in the previous
subsection) must be added to $G$ as an external perturbation. A
bifurcation of a rotationally-asymmetric solution implies a change
of sign of $\delta G/\delta c$, meaning that the rotationally
symmetric one has become unstable to that particular perturbation.
Hence, the instability-onset line for perturbations that break
rotational symmetry numerically coincides with the phase boundary
between the ``gas'' and the ``double cluster'' phases, where
$\text{Hes}_{(E,L)}[S]=0$, displayed in the phase diagram.


\section{Outlook and final remarks}

We have presented an analysis of the equilibrium properties of a
self-gravitating and rotating gas using a microcanonical
mean-field approach. Our main result concerns the spontaneous
breaking of the rotational symmetry, which takes place at high
angular momentum and gives rise to non-trivial density profiles,
e.g. ``double stars'': a rapidly rotating $N$-body system kept
together by gravitation only at equilibrium spontaneously
organizes in two distinct dense clusters, provided the ratio
between rotational and gravitational energy is sufficiently high
and the total energy is not too large. We have derived the global
phase diagram of the model and discussed the related thermodynamic
picture, providing a phenomenological description of the phase
transitions occurring and analyzing the stability of high-energy
rotationally-symmetric equilibrium states against perturbations
that break rotational symmetry, showing that non-trivial
rotationally symmetric solutions such as rings become unstable in
the mixed (phase coexistence) region. To our knowledge, these
results constitute the most complete equilibrium description of a
self-gravitating and rotating system to date. To conclude, we
would like to put forward some final remarks and open problems.
\begin{enumerate}
\item[i.] While it is certainly possible to improve on the results
presented here by increasing the maximum order of the even
harmonics included in the calculation, we do not expect any major
qualitative difference with the picture we describe here
($l_{\text{max}}=16$). The inclusion of odd harmonics would
instead lead to the formation of asymmetric double clusters, and
to a more complete phase diagram and a full classification of the
different possible equilibrium configurations as a consequence.
This issue will be treated elsewhere \cite{prep}.
\item[ii.] Our results stem from a mean-field analysis, in which
particle-particle correlations are completely neglected.
\item[iii.] We have not studied how our results depend on $\Theta$,
namely on the average density of the system. We mentioned that
this is an important issue, certainly worth to be investigated
with great care.
\item[iv.] A general stability analysis requires, as we said, the
study of the eigenvalue problem for $K$, Eq. (\ref{ka}). The most
interesting open question is that of marginal stability, that is
solutions of (\ref{eigen}) with zero eigenvalue. However, the
approach we discussed in the last section, connecting the
stability analysis to the bifurcation analysis, describes
correctly the stability of the different states against
perturbations that break rotational symmetry (within the included
number of harmonics).
\item[v.] From a physical point of view, it is known that energy
and angular momentum are possibly not the only conserved
quantities to be taken into account if one wants to recover some
observational features of e.g. galaxies \cite{contopoulos60}.
\item[vi.] Of course, we have dealt with equilibrium properties
exclusively, and provided a kind of classification of the
different possible states of the system. This clearly leaves open
many important questions concerning dynamics and relaxation
mechanisms, which are believed to be particularly subtle in
self-gravitating systems \cite{lyndenbell67}.
\end{enumerate}

On a more general level, the equilibrium properties of
non-extensive Hamiltonian systems (self-gravitating and rotating
systems being the most important example) can be well described by
Boltzmann's principle (\ref{duino}). We would finally like to
stress the potentialities of the method presented here for
studying the equilibrium properties of systems with long-range
(see footnote 1) forces. Its basic ingredients are: (a)
microcanonical statistics and (b) a series expansion of the
potential using a proper basis set. This same scheme can easily be
modified to work with other types of long-range potentials (e.g.
Coulomb or Yukawa). In the light of this last remark,
self-gravitating and rotating systems (for which the use of
``corrective'' techniques such as the Kac-Uhlenbeck-Hemmer method
\cite{kac63} is substantially not helpful in clarifying its
complex thermodynamic structure) should be considered as a
fundamental testing ground for techniques to analyze the
statistical equilibrium properties of systems with long-range
interactions.

\bigskip

\textbf{Acknowledgments.} We wish to thank P.H. Chavanis and O.
Fliegans for important comments, and the Deutsche
Forschungsgemeinschaft (DFG) for financial support. We also
acknowledge useful discussions with the participants of the
conference on ``Dynamics and Thermodynamics of Systems with Long
Range Interactions'' (Les Houches, February 2002).

\end{document}